\documentclass[conference]{IEEEtran}
\usepackage[mathcal]{euscript}
\usepackage{amsmath}
\usepackage{amsthm}
\usepackage{amssymb}
\usepackage{graphicx}
\usepackage{algorithmic}
\usepackage{algorithm}
\usepackage{epsfig}
\usepackage{booktabs}% http://ctan.org/pkg/booktabs
\usepackage{colortbl}% http://ctan.org/pkg/colortbl
\usepackage{amsmath}% http://ctan.org/pkg/amsmath
\usepackage{xcolor}% http://ctan.org/pkg/xcolor
\usepackage{graphicx}% http://ctan.org/pkg/graphicx
\usepackage{cite} % aggregate refrences
\usepackage{etex}
\usepackage{tikz}
\usepackage{tikz-3dplot}
\usepackage{pst-3dplot}
\usepackage{subfigure}

\usetikzlibrary{positioning}

\providecommand{\mbf}[1]{\mathbf{#1}}						 % Math boldface
\providecommand{\wt}[1]{\widetilde{#1}}					 % Wide tilde
\providecommand{\mbfwt}[1]{\wt{\mathbf{#1}}}				 % Math boldface with wide tilde
\providecommand{\mc}[1]{\mathcal{#1}}				 % Math Cal
\providecommand{\bsym}[1]{\boldsymbol{#1}}				 % Bold symbols by the bm package. Better than 																					 amsmath \pmb if the font exists.
\providecommand{\bsymwt}[1]{\widetilde{\boldsymbol{#1}}}	 % Bold symbol with wide tilde

\IEEEoverridecommandlockouts

\begin{document}

\title{Three-dimensional (3D) Channel Modeling between Seaborne MIMO Radar and MIMO Cellular System}

\author{\begin{tabular}{c}
 Awais Khawar, Ahmed Abdelhadi, and T. Charles Clancy \\
\{awais, aabdelhadi, tcc\}@vt.edu \\
Ted and Karyn Hume Center for National Security and Technology\\
Bradley Department of Electrical and Computer Engineering\\
Virginia Tech, Arlington, VA, 22203, USA
\end{tabular}
\thanks{This work was supported by DARPA under the SSPARC program. Contract Award Number: HR0011-14-C-0027. The views, opinions, and/or findings contained in this article are those of the authors and should not be interpreted as representing the official views or policies of the Department of Defense or the U.S. Government.

Approved for Public Release, Distribution Unlimited

}

}

\maketitle

\begin{abstract}
Sharing radar spectrum with communication systems is an emerging area of research. Deploying commercial wireless communication services in radar bands give wireless operators the much needed additional spectrum to meet the growing bandwidth demands. However, to enable spectrum sharing between these two fundamentally different systems interference concerns must be addressed. In order to assess interference concerns we design a three-dimensional (3D) channel model between radar and cellular base stations (BSs) in which the radar uses a two-dimensional (2D) antenna array and the BS uses a one-dimensional (1D) antenna array. We formulate a line-of-sight (LoS) channel and then propose an algorithm that mitigates radar interference to BSs. We extend the previously proposed null space projection algorithm for 2D channels to 3D channels and show that effective nulls can be placed by utilizing both the azimuth and elevation angle information of BSs. This results in effective interference mitigation. In addition we show that the 3D channel model allows us to accurately classify the size of radar's search space when null space projection algorithm is used for interference mitigation.

%
%In this letter, we propose   
%
%
% as it promises vast amount of spectrum that was available for exclusive radar use for commercial operations.
%
%
%In this paper, we model a 3D channel between a seaborne MIMO radar and an on-shore cellular system. A 2D antenna array is introduced at the seaborne MIMO radar which gives 
%

\end{abstract}


\section{Introduction}

Today's wireless communication operators urgently require more spectrum to increase capacity and meet consumer demands. A classic practice to free up more usable spectrum for wireless systems has been to relocate incumbents to some other band. However, not only this process is time consuming it is also very expensive for incumbents to relocate. An emerging way forward, to avoid the above mentioned issues, is to share spectrum across government agencies and commercial services. In the United States, the Federal Communications Commission (FCC) is considering to allow commercial wireless operation in the 3.5 GHz radar band \cite{FCC12_SmallCells}. This initiative has the potential to leverage new methods of spectrum sharing among heterogeneous systems such as small cells, radars, and satellite systems. However, a fundamental limitation for spectrum sharing is interference caused to coexisting systems. Novel interference mitigation methods are required for coexistence of these fundamentally different systems. This letter addresses this issue by modeling a 3D channel between radar and communication systems and proposes an algorithm that mitigates radar interference to cellular BSs. 

%mitigation studies the impact of a proposed interference mitigation scheme.

A fundamental area of research in wireless communications is channel modeling as it allows performance evaluation of transmission techniques. 
Traditionally, wireless channel models have been designed to capture the azimuth i.e. 2D channels. A comprehensive introduction to 2D wireless channel modeling, propagation modeling, and statistical description of channels can be found in \cite{Mol10} and references there in. Recently, there has been a surge in modeling wireless channels that capture both the azimuth and elevation directions known as 3D channels. 3D channels allow very precise beamforming for users on ground and in buildings. As a result the wireless research community is heavily involved in designing and standardizing 3D channel models that can reap the benefits of both the azimuth and elevation beamforming \cite{KKA14}. However, the efforts so far have been limited to model 3D channels between wireless communication systems and to the best of our knowledge no work exists on 3D channel modeling between wireless communication and radar systems. A 3D channel model can enable radar systems to place accurate nulls in azimuth and elevation angles of BSs in order to mitigate radar interference. Therefore, in order to study performance evaluation and interference mitigation techniques in a spectrum sharing scenario, the problem of 3D channel modeling is of prime importance.

In this letter, we formulate a 3D MIMO channel model between MIMO radar and MIMO cellular system. We consider a 2D antenna array at the radar and a 1D array at the BS. Using our proposed channel model we demonstrate the efficacy of a novel null space projection algorithm in 2D which mitigates radar interference. The rest of this letter is organized as follows. Section \ref{sec:models} briefly presents MIMO radar architecture and spectrum sharing scenario. Section \ref{sec:channel} models 3D MIMO channel between seaborne radar and cellular system. Section \ref{sec:2DNSP} proposes a novel interference mitigation algorithm for 3D channels. Section \ref{sec:search} discuss the volume illuminated by spectrum sharing MIMO radars that are subject to interference mitigation. Section \ref{sec:sim} discusses the simulation results. Section \ref{sec:conc} concludes the letter.

\section{Spectral-Coexistence Models}\label{sec:models}
In this section, we briefly introduce the fundamentals of MIMO radar and discuss our radar-communication system spectrum sharing scenario.

\subsection{MIMO Radar}
   
We consider a shipborne mono-static MIMO radar with $M= M_h \times M_v$ colocated antennas in a uniform rectangular array (URA), where $M_h$ is the number of horizontal antenna elements in a given row and $M_v$ is the number of vertical antenna elements in a given column. We assume that the inter-element distance between adjacent antenna elements on any given column and row is $d_s$. The target is assumed to be a point target %and its orientation with respect to radar is shown in Figure \ref{fig:orientation} 
which has azimuth and elevation angles represented by $\theta$ and $\phi$, respectively.  
%We define the transmit steering matrix as
%\begin{equation}
%\mbf A(\theta,\phi) \triangleq \mbf u(\theta,\phi) \mbf v^T(\theta,\phi)
%\end{equation}
%where    
%\begin{equation}
%   \mbf u(\theta,\phi) = \begin{bmatrix}
%   1 &e^{j2\pi d_s \sin \theta \cos \phi} &\cdots & e^{j2\pi (M_h-1)d_s \sin \theta \cos \phi}
%   \end{bmatrix}
%\end{equation}   
% and   
%\begin{equation}
%   \mbf v(\theta,\phi) = \begin{bmatrix}
%   1 &e^{j2\pi d_s \sin \theta \sin \phi} &\cdots & e^{j2\pi (M_h-1)d_s \sin \theta \sin \phi}
%   \end{bmatrix}
%   \end{equation} 
%Azimuth and elevation angles are represented by $\phi$ and $\theta$, respectively.   
  
Let $\mbf x$ be the transmitted waveform vector which 
%satisfies
%= x_1(t), \ldots, x_K(t)$ be the $K \times 1$ transmitted waveform vector which satisfies 
%\begin{equation}
%   \mbf x(t) = \begin{bmatrix}
%   &x_1(t), &\ldots, &x_K(t)
%   \end{bmatrix}
%   \end{equation}   
%        
%We are interested in 
simultaneously illuminates the target area of interest and nulls interference to BSs present in a 2D spatial sector.

\subsection{Spectral Coexistence Scenario}
We consider a spectrum sharing scenario in which a cellular system is deployed in the radar band. Each BS of the cellular system is equipped with $N$ transmit and receive antennas in a uniform linear array (ULA). Each BS antenna is at a specific azimuth and elevation angle with respect to radar. We assume a Frequency Division Duplex (FDD) cellular network in which BSs are operating in the radar band and UEs are operating at non-radar frequencies. Therefore, we focus on interference caused by the radar operation to the base stations only. We consider a single cell or a BS with $\mc K$ users, without loss of generality, that receives the following signal in the presence of radar
\begin{equation}
{\mbf{r}}=\sum_{i=1}^{\mc K} \mbf G_i \mbf s_i + {\mbf{H}} \mbf x + \mbf{n} 
\end{equation}  
where $\mbf G_i$ is the channel gain between the BS and the $i^{\text{th}}$ user, $\mbf H$ is the channel gain between the BS and the radar, and $\mbf x$ is the interfering signal from the MIMO radar. This interference from radar can be mitigated by projecting radar waveform onto null space of interference channel between radar and BS \cite{S.SodagariDec.2012, GhorbanzadehMilcom2014, KAC14_TWS, KAC14_TDetect, Channel2D, ChannelLOS, KAC14_QPSK, A.Khawar, KAC14_Milcom, KAC14DySPANWaveform, KAC14ICC, KAC+14ICNC, SAC+15, SKA+14DySPAN}. Previous work considers null space projection for 2D channels \cite{Channel2D} which is extended in this paper to 3D channels.  
%which we want to mitigate by designing a channel model and then projecting radar signal onto null space of the channel . 

\begin{figure}[t]
    \centering
     \includegraphics[trim=0.5in 1.3in 1in 1in, width=3.4in]{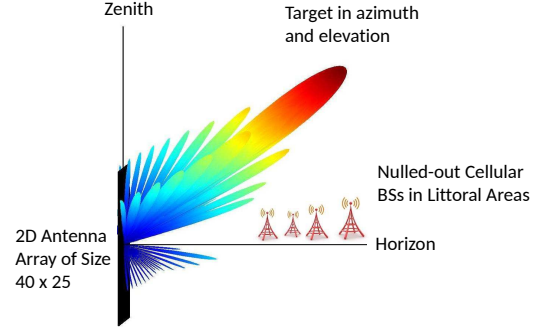} 
    \caption{Spectrum sharing scenario between radar and cellular BSs. The portion containing BSs is nulled-out to protect it from the radar's interference. 
%A view of radar's 3D beampattern along with communication systems. 
} 
   \end{figure}

\section{3D Channel Modeling}\label{sec:channel}

In this letter, we formulate a 3D channel model between radar and communication system. We introduce a 2D uniform rectangular array (URA) antenna array at the radar and a 1D uniform linear array (ULA) at the BS.  We assume $M = M_h \times M_v$ antenna elements, where $M_h$ antenna elements exist in the horizontal/row direction and $M_v$ antenna elements exist in the vertical/column direction with equal element spacing $d_s$. Each element of the 2D antenna array can be labeled by $s$, where $s \triangleq (l,k) = (k-1)M_v+l$, for $k=1,2,\ldots,M_h$ and $l=1,2,\ldots,M_v$.

We proceed by deriving 3D channel model for line-of-sight (LoS) case. In the absence of any beamforming, the 3D channel characteristics are dependent on element radiation pattern, antenna spacing, and propagation of each ray path. In the LoS case, for the two adjacent vertical antenna elements $s$ and $s+M_v$ from the same column, the channel coefficients between receive-transmit antenna pairs $(u,s)$ and $(u,s+M_v)$, i.e., $H_{u,s}$ and $H_{u,s+M_v}$, have the following property \cite{JHL13}
\begin{equation}\label{eqn:shift}
H_{u,s+M_v} \approx H_{u,s} e^{-\jmath 2\pi/\lambda d_s \cos (\phi)}
\end{equation}
i.e. the difference between the two adjacent elements can be captured by the phase shift due to propagation. In equation \eqref{eqn:shift}, $\phi$ is the angle of departure in elevation domain.

For a BS with $N$ antenna elements in a uniform linear array (ULA), the $N \times M \:$%_h \times M_v$ 
MIMO channel is partitioned into $M_h$ sub-matrices given by
\begin{equation}
\mbf H_h =\begin{bmatrix}
\mbfwt H_{k=1} &\mbfwt H_{k=2} &\cdots &\mbfwt H_{k=M_h}
\end{bmatrix}
\end{equation}
where each $\mbfwt H_k$ has size $N \times M_v$ and is given as
\begin{align*}
&\mbfwt H_{k}= \\
&\begin{bmatrix} H_{1,(k-1)M_v+1} & H_{1,(k-1)M_v+2} &\ldots & H_{1,(k-1)M_v+M_v} \\
H_{2,(k-1)M_v+1} & H_{2,(k-1)M_v+2} & \ldots & H_{2,(k-1)M_v+M_v} \\
\ldots & \ldots &\ldots &\ldots \\
H_{N,(k-1)M_v+1} & H_{N,(k-1)M_v+2} & \ldots & H_{N,(k-1)M_v+M_v}
\end{bmatrix}
\end{align*}
for $k=1,2,\ldots, M_h$.

Using equation \eqref{eqn:shift} we can write the above channel matrix as
\begin{equation}
\mbf H_h = \begin{bmatrix}
\mbfwt H_{k=1} &\mbfwt H_{k=2} &\cdots &\mbfwt H_{k=M_h}
\end{bmatrix} = \mbf a_h(\theta) \otimes \mbfwt H_{k=1} 
\end{equation}
where $\otimes$ is the Kronecker product and
\begin{equation*}
\mbf a_h (\theta) = \begin{bmatrix}
1 &e^{-\jmath 2\pi/\lambda d_s \cos (\theta)} &\cdots &e^{-\jmath 2\pi/\lambda (M_h-1) d_s \cos (\theta)}
\end{bmatrix}
\end{equation*}
is the steering vector in azimuth domain. Using the same arguments we can write the partitioned channel model in elevation domain as
\begin{equation}
\mbf H_v = \begin{bmatrix}
\mbfwt H_{l=1} &\mbfwt H_{l=2} &\cdots &\mbfwt H_{l=M_v}
\end{bmatrix}
\end{equation}
and each $\mbfwt H_l$ has size $N \times M_v$. Using equation \eqref{eqn:shift} we can write above as
\begin{equation}
\mbf H_v = \begin{bmatrix}
\mbfwt H_{l=1} &\mbfwt H_{l=2} &\cdots &\mbfwt H_{l=M_v}
\end{bmatrix} = \mbf a_v(\phi) \otimes \mbfwt H_{l=1} 
\end{equation}
where
\begin{equation*}
\mbf a_v(\phi) = \begin{bmatrix}
1 &e^{-\jmath 2\pi/\lambda d_s \cos (\phi)} &\cdots &e^{-\jmath 2\pi/\lambda (M_v-1) d_s \cos (\phi)}
\end{bmatrix}
\end{equation*}
is the steering vector in elevation domain.

\section{2D Null Space Projection}\label{sec:2DNSP}

In this section, we introduce NSP in both azimuth and elevation domains by considering the 3D channel model designed in the previous section. %We proceed by projecting radar signal first into azimuth domain, i.e.,
%\begin{equation}
%\mbfwt x_h(t) = \mbf P_h \mbf x(t)
%\end{equation}
%where $P_h$ is the projection matrix in the horizontal or azimuth domain. Similarly, the projection in elevation domain is given by
%\begin{equation}
%\mbfwt x_v(t) = \mbf P_v \mbf x(t)
%\end{equation}
%The covariance matrix of NSP transmitted waveform is given by
%\begin{equation}
% \mbf R_{NSP} = \mbf R_{h,NSP} + \mbf R_{v,NSP}
% \end{equation} 
%where
%\begin{align}
%\mbf R_{h,NSP} &= \int_{T_o} \mbfwt x_h(t) \mbfwt x_h^H(t) dt \\
%\mbf R_{v,NSP} &= \int_{T_o} \mbfwt x_v(t) \mbfwt x_v^H(t) dt 
%\end{align}
%The 3D beampattern of NSP signal is given as
%\begin{equation}
%G(\theta,\phi) = |\mbf a_v^T(\phi) \: \mbf R_{NSP} \: \mbf a_h(\theta)|
%\end{equation}
%
%\begin{equation}
%\mbf R_{Elv} = \mbf R \mbf a_h \mbf a_v^T * \mbf R_{Azm}
%\end{equation}
We define projection algorithm which projects radar signal onto null space of 3D interference channel, i.e., in elevation $\mbf H_{v}$ and azimuth $\mbf H_{h}$. Assuming, the MIMO radar has channel state information, we can perform singular value decomposition (SVD) in both domains to find the null space and then construct a projector matrix. We proceed by first finding SVD of $\mbf H_{v}$, i.e., 
\begin{equation}
\mbf H_{ v} = \mbf U_{v} \bsym \Sigma_{v} \mbf V_{v}^H.
\end{equation}
Now, let us define 
\begin{equation}
\bsymwt \Sigma_{v} \triangleq \text{diag} (\wt \sigma_{v,1}, \wt \sigma_{v,2}, \ldots, \wt \sigma_{v,p})
\end{equation}
where $p \triangleq \min (N,M_v)$ and 
$\wt \sigma_{v,1} > \wt \sigma_{v,2} > \cdots > \wt \sigma_{v,q} > \wt \sigma_{v,q+1} = \wt \sigma_{v,q+2} = \cdots = \wt \sigma_{v,p} = 0$ are the singular values of $\mbf H_{v}$. Next, we define
\begin{equation}
\bsymwt {\Sigma}_{v}^\prime \triangleq \text{diag} (\wt \sigma_{v,1}^\prime,\wt \sigma_{v,2}^\prime, \ldots, \wt \sigma_{v,M_v}^\prime)
\end{equation}
where
\begin{align}
\wt \sigma_{v,u}^\prime \triangleq
\begin{cases}
0, \quad \text{for} \; u \leq q,\\
1, \quad \text{for} \; u > q.
\end{cases}
\end{align}
Using above definitions we can now define our projection matrix in elevation, i.e.,
\begin{equation}\label{eqn:ProjDefinition}
\mbf P_{v} \triangleq \mbf V_{v} \bsymwt \Sigma_{v}^\prime \mbf V_{v}^H.
\end{equation}
Using similar arguments we can define projection matrix in azimuth, i.e.,
\begin{equation}
\mbf P_{h} \triangleq \mbf V_{ h} \bsymwt \Sigma_{h}^\prime \mbf V_{ h}^H.
\end{equation}
Using the above defined projection matrices we can write covariance matrices of projected waveform in elevation as
\begin{equation}
\mbf R_{Elv}^{Null} = \mbf P_{v} \mbf R_{Elv} \mbf P_{v}^T
\end{equation}
and azimuth as
\begin{equation}
\mbf R_{Azm}^{Null} = \mbf P_{h} \mbf R_{Azm} \mbf P_{h}^T.
\end{equation}
The covariance matrix of NSP transmitted waveform is given by
\begin{equation}
 \mbf R_{NSP} = \mbf R_{v,NSP} + \mbf R_{h,NSP}
 \end{equation} 
where
\begin{align}
\mbf R_{v,NSP} &= \mbf R_{Azm}^{Null} \mbf a_h(\theta) \mbf a_v^T(\phi) \mbf R_{Elv} \\
\mbf R_{h,NSP} &= \mbf R_{Azm} \mbf a_h(\theta) \mbf a_v^T(\phi) \mbf R_{Elv}^{Null}
\end{align}
where $\mbf R_{Elv}$ and $\mbf R_{Azm}$ are waveform covariance matrices in elevation and azimuth, respectively. 
%where
%\begin{align}
%\mbf R_{Azm}^{Null} &=  \mbf P_v \mbf R_{Azm} \mbf P_v^T \\
%\mbf R_{Elv}^{Null} &= \mbf P_h \mbf R_{Elv} \mbf P_h^T.
%\end{align}
%\begin{align}
%\mbf P_v &=   \\
%\mbf P_h &= 
%\end{align}
%
%\begin{equation}
% \mbf R_{NSP} = \mbf R_{h,NSP} + \mbf R_{v,NSP}
% \end{equation} 
%where
%\begin{align}
%\mbf R_{h,NSP} &= \int_{T_o} \mbfwt x_h(t) \mbfwt x_h^H(t) dt \\
%\mbf R_{v,NSP} &= \int_{T_o} \mbfwt x_v(t) \mbfwt x_v^H(t) dt 
%\end{align}
The 3D beampattern of NSP signal is given as
\begin{equation}
G(\theta,\phi) = |\mbf a_v^T(\phi) \: \mbf R_{NSP} \: \mbf a_h(\theta)|.
\end{equation}

%\begin{equation}
%\mbf R_{Elv} = \mbf R \mbf a_h \mbf a_v^T * \mbf R_{Azm}
%\end{equation}

\section{Radar Search Space}\label{sec:search}

Radar search space is normally specified by a search solid angle $\Omega$ in steradians. If we define the radar search volume extent for both azimuth and elevation, in degrees, as $\theta$ and $\phi$, the search volume can be given as \cite{POMR}
\begin{equation}
 \Omega = \frac{\theta \: \phi}{(57.296)^2} \quad \text{(steradians)}.
 \end{equation} 
Let $\Omega_{\text{Null}}$ denote the sector, azimuth and elevation angles, of cellular BSs that needs to be  nulled to protect BSs from radar interference. Then, due to NSP the search volume, shown in Figure 2, can be given as
\begin{equation}
 \Omega_{\text{NSP}} = \frac{\theta \phi - \Omega_{\text{Null}}}{(57.296)^2} \quad \text{(steradians)}
%\frac{\theta \: ( \phi - |\phi_{\text{BS}}| )}{(57.296)^2} \quad \text{(steradians)}.
\end{equation} 
%%%%%%%%%%%%%%%%%%%%%%%%%%%%%%%%%%%%%%%%%%%%%%%%
%% Sphere with blocked out Region %%%%%%%%%%%%%%
%%%%%%%%%%%%%%%%%%%%%%%%%%%%%%%%%%%%%%%%%%%%%%%%

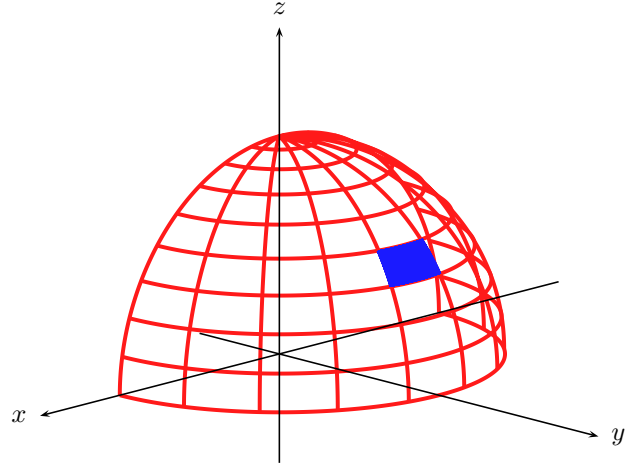
\begin{figure}\label{fig:volume}
\centering
\begin{pspicture}(0,0)(0,5)
\psset{unit=3cm,Alpha=45, Beta=15,linecolor=red!90,linewidth=1.5pt}
\parametricplotThreeD[xPlotpoints=500, yPlotpoints=10](0,180)(0,90){
    /r 1.0 def
    r t cos mul u sin mul 
    r t sin mul u sin mul 
    r u cos  mul }
\parametricplotThreeD[xPlotpoints=500, yPlotpoints=10](0,90)(0,180){
    /r 1.0 def
    r u cos mul t sin mul 
    r u sin mul t sin mul 
    r t cos  mul }
\psset{unit=3cm,Alpha=45, Beta=15,linecolor=blue!90,linewidth=1.5pt}
    \parametricplotThreeD[xPlotpoints=500, yPlotpoints=1000](50,60)(80,100){ %use it for blockfading, slow, so use it for final version
%use it for quick plots
    /r 1.0 def
    r u cos mul t sin mul 
    r u sin mul t sin mul 
    r t cos  mul }
\pstThreeDCoor[xMin=-1.75,xMax=1.5,yMin=-.5,yMax=2.0, zMin=-.5,zMax=1.5,linecolor=black,linewidth=0.6pt]
%\caption{hjh}
\end{pspicture}
\qquad
\qquad
\vspace{1.2cm}
\caption{Search space of spectrum sharing MIMO radar. The part containing cellular BSs is not part of the radar's illuminated area due to the nulls placed to protect BSs from its interference. That is why the blocked-out part (blue part) of the figure is an illustration of the nulls placed in that particular azimuth-elevation sector.}
\end{figure}

\section{Numerical Examples} \label{sec:sim}
In this section, we provide numerical examples of 3D channel modeling between radar and communication system and its applications using null space projection. For the simulation setup we assume that the radar has a $40 \times 25$-element URA and the BS has a $10$-element ULA. We assume a target is located at $\theta =0^\circ$ and $\phi=50^\circ$ and communication systems are located in a sector of azimuth angles $\theta =-45^\circ$ to $-40^\circ$ and elevation angles $\phi=5^\circ$ to $15^\circ$. We intend to null this sector to mitigate radar interference to this area. We provide two figures of the same example with different views to better understand the achieved results.

\begin{figure}
\centering
\includegraphics[width=\linewidth]{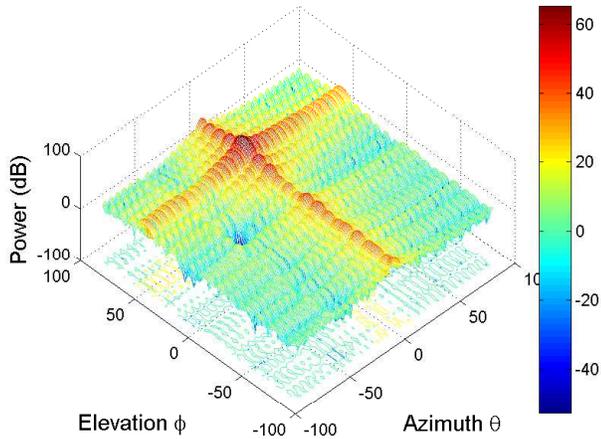} 
	\caption{A 3D view of interference mitigation at BSs. Target is located at $\theta =0^\circ$ and $\phi=50^\circ$ and BSs are located in a sector of azimuth angles $\theta =-45^\circ$ to $-40^\circ$ and $\phi=5^\circ$ to $15^\circ$ and we intend to null this sector to mitigate radar interference to this area.}
		\label{fig:3D}
\end{figure}

\begin{figure}
\centering
\includegraphics[width=\linewidth]{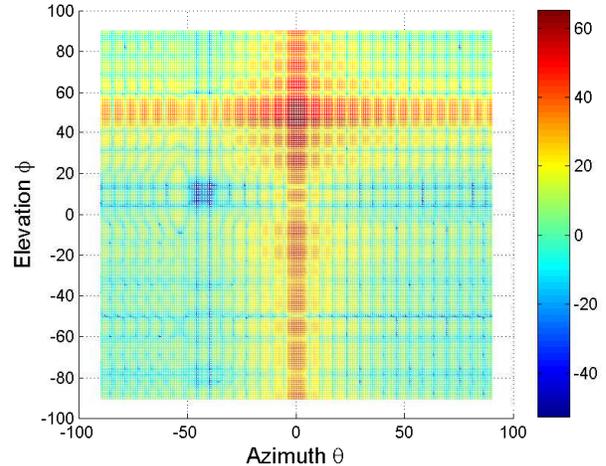} 
	\caption{An image of interference mitigation at BSs. The intensity/magnitude of radar signal can be observed at the target and BSs. Note the difference between the power received at target vs. the power received at BSs.}
		\label{fig:image}
\end{figure}

In Figures \ref{fig:3D} and \ref{fig:image}, the impact of 2D NSP is studied on the radar beampattern. It can be noted that the target can be accurately localized when it is far from the nulled areas. Moreover, the power received at the target is 60 dB which is significantly higher than the power received at the BSs which is less than -40 dB in this example. Thus, the proposed approach satisfies both the radar mission objective of target detection and  simultaneously mitigates interference to BSs. Moreover, only the areas where BSs are present are nulled.

\begin{figure}
\centering
\includegraphics[width=\linewidth]{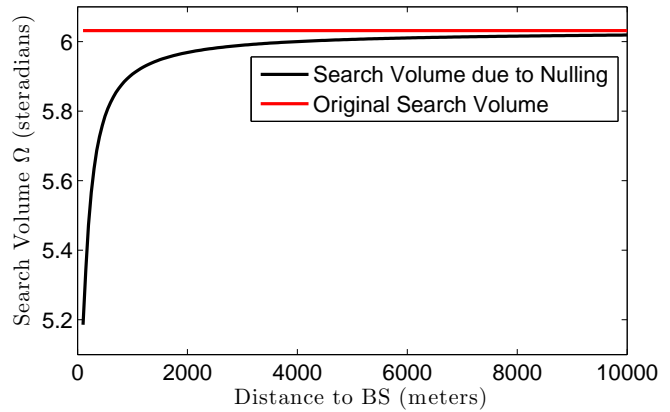} 
	\caption[Search volume when radar is in deep sea]{Search volume for radar located far from BSs. Search-able space is 180$^\circ$ in azimuth, i.e., from -90$^\circ$ to 90$^\circ$ and 110$^\circ$ in elevation, i.e, from -20$^\circ$ to 90$^\circ$.  So the searchable volume in percentage at a distance of: 500 m is 95.9\%, 2 km is 99\%, and at 8 km is 99.8\%. At longer distances the search volume with and without NSP is very close.}
		\label{fig:VolvsDist_Comparison}
\end{figure}

In Figure \ref{fig:VolvsDist_Comparison}, we show the volume illuminated by the spectrum sharing radar as a function of its distance from nulled BSs. We assume the search-able space is 180$^\circ$ in azimuth, i.e., from -90$^\circ$ to 90$^\circ$ and 110$^\circ$ in elevation, i.e, from -20$^\circ$ to 90$^\circ$. Note that for close-up range the difference in search volume with NSP and without NSP is significant, i.e., for ships docket at the harbor or close to shore line will experience some degradation in search volume due to NSP. However, if the radar is more than 2000 meters away from BSs the loss in search space is minimal.

\section{Conclusion}\label{sec:conc}
Spectrum sharing among heterogeneous systems is a promising solution to solve the spectrum congestion problem. In this letter, we explored a spectrum sharing scenario between seaborne radar and on-shore cellular system. We formulated a 3D channel model between radar and communication system. We assumed a 2D antenna array at the radar and a 1D antenna array at the BS. In addition, we introduced a new interference mitigation algorithm that projected radar waveform onto null space of 3D channel to mitigate radar interference to cellular BSs. The efficacy of the proposed algorithm was shown using simulation results. Search space of spectrum sharing MIMO radars using NSP was also explored and a relation to quantify the size of search space was provided. It was shown that the reduction in search space was directly related to the areas nulled to protect BSs.

\bibliographystyle{ieeetr}
\bibliography{IEEEabrv,channel3D}
\end{document}